\documentclass[lettersize,journal]{IEEEtran}
\IEEEoverridecommandlockouts
\usepackage{cite}
\usepackage{amsmath,amssymb,amsfonts}
\usepackage{algorithmic}
\usepackage{graphicx}
\usepackage{textcomp}
\usepackage{gensymb}
\usepackage{xcolor}
\usepackage{booktabs}
\usepackage{graphicx}
\usepackage{verbatim}
\usepackage{adjustbox}
\usepackage{url}
 \usepackage{array}
 \usepackage{pdflscape}
 \usepackage{longtable}
\usepackage{caption}
\usepackage{siunitx}
\def\BibTeX{{\rm B\kern-.05em{\sc i\kern-.025em b}\kern-.08em
    T\kern-.1667em\lower.7ex\hbox{E}\kern-.125emX}}
\begin{document}

\title{CKMImageNet{:} A Dataset for AI-Based Channel Knowledge Map Towards Environment-Aware Communication and Sensing \\

\thanks{Part of this work has been presented at the IEEE ICCC Workshops 2024~\cite{ckmimagenet}.}
}

\author{Zijian~Wu, Di~Wu, Shen~Fu, Yuelong~Qiu, and 
Yong~Zeng,~\IEEEmembership{Fellow,~IEEE}
\thanks{This work was supported by the Natural Science Foundation for Distinguished Young Scholars of Jiangsu Province with grant number BK20240070.}
\thanks{This research work is supported by the Big Data Computing Center of Southeast University.}
\thanks{The authors are with the National Mobile Communications Research Laboratory, Southeast University, Nanjing 210096, China.
Di Wu and Yong Zeng are also with the Purple Mountain Laboratories, Nanjing 211111, China (e-mail: {\{wuzijian, studywudi, sfu, yl\_qiu, yong\_zeng\}}@seu.edu.cn). (\textit{Corresponding author: Yong Zeng.})}
}

\maketitle

\begin{abstract}
 With the increasing demand for real-time channel state information (CSI) in sixth-generation (6G) mobile communication networks, channel knowledge map (CKM) emerges as a promising technique, offering a site-specific database that enables environment-awareness and significantly enhances communication and sensing performance by leveraging a priori wireless channel knowledge. However, efficient construction and utilization of CKMs require high-quality, massive, and location-specific channel knowledge data that accurately reflects the real-world environments. Inspired by the great success of ImageNet dataset in advancing computer vision and image undertanding in artificial intelligence (AI) community, we introduce CKMImageNet, a dataset developed to bridge AI and environment-aware wireless communications and sensing by integrating location-specific channel knowledge data, high-fidelity environmental maps, and their visual representations. CKMImageNet supports a wide range of AI-driven approaches for CKM construction with spatially consistent and location-specific channel knowledge data, including both supervised and unsupervised, as well as discriminative and generative AI methods. The dataset is built using advanced ray-tracing techniques, ensuring high fidelity and environmental accuracy. By addressing key challenges in CKM construction and enabling AI models to learn environment-aware propagation patterns, CKMImageNet may serve as a foundational tool for advancing environment-aware 6G systems, ranging from network planning such as communication base station (BS) site selection and sensing anchor node placement, to pro-active resource allocation such as beam alignment, power allocation, interference avoidance, clutter rejection, and robot trajectory planning. Compared with existing datasets like RadioMapSeer, CKMImageNet not only provides numerical and visual representation to channel gain values, but also more diversified channel knowledge like multipath angle of arrivals (AoAs) and path delays. Moreover, the dataset offers images with multiple sizes to cater to different application scenarios.

\end{abstract}


\section{Introduction}
The paradigm shift from the conventional environment-unaware towards future environment-aware communications and sensing is mainly driven by the growing challenge of acquiring real-time channel state information (CSI) for all wireless links~\cite{ckm}. Specifically, with denser wireless infrastructures and larger-dimensional channels\cite{di2014, you2021towards}, traditional CSI acquisition methods that heavily rely on real-time channel training becomes costly or even infeasible. On the other hand, the sixth-generation (6G) mobile communication networks are expected to benefit from a multitude of data sources containing high-quality location-tagged channel data, facilitated by advancements in positioning and sensing technologies, together with artificial intelligence (AI) for powerful data mining. This abundance of data enables better perception of the local wireless environment, presenting a new opportunity to address the CSI acquisition challenge that is closely coupled with the local wireless environment. 
This has motivated the recent development of channel knowledge map (CKM) as a promising new solution to enable 6G environment-aware wireless communications and sensing\cite{zeng2021toward}.

CKM is a site-specific database that provides location-specific channel knowledge \cite{zeng2021toward}. It offers a promising method to enable environment-awareness, which facilitates or even avoid sophisticated real-time CSI acquisition or redundant environment sensing. Different from conventional concept like radio environment map (REM)\cite{zhao2006network}, CKM directly reflects the intrinsic wireless channel properties of the local environment, offering a range of channel knowledge such as region-specific channel modeling parameters, and location-specific large-scale or small-scale channel knowledge such as channel gains, multi-path angle of arrivals/departures (AoAs/AoDs), propagation delays, etc.  

CKM is expected to play an important role in various communication and sensing tasks. For example, for communication tasks such as environment-aware site planning, beamforming, channel prediction, and resource allocation, CKM can enhance their performance by offering a prior channel knowledge. For millimeter-wave (mmWave) systems, two particular types of CKMs, namely  channel path map (CPM) and beam-index map (BIM) have been proposed to enable training-free analog or hybrid analog/digital beamforming \cite{wu2023environment1}. 
In\cite{wu2023environment2}, multi-point collaborative beamforming scheme was researched using channel angle map (CAM), by considering both scenarios with known and unknown user locations. 
CKM has been also leveraged for environment-aware active and passive beamforming for reconfigurable intelligent surface (RIS)-assisted wireless communications\cite{taghavi2023environment}. For channel prediction, CKM provides prior knowledge like channel correlation matrices, enabling low-overhead pilot designs\cite{qiu2024ckm, jiang2025interference}. 
Combining CKM with historical user CSI, an adaptive low-complexity channel estimator was proposed in\cite{wang2024channel}. 
Furthermore, the authors in\cite{wu2024environment} decomposed the wireless environment into quasi-static and dynamic components, and proposed an efficient channel estimation method leveraging CKM and dynamic sensing information. 
Furthermore, CKM has also been utilized to accomplish tasks such as unmanned aerial vehicle (UAV) trajectory planning and signal transmission rate scheduling\cite{zhan2023aerial, yue2024channel}.
In addition, CKM advances sensing and localization capabilities through location-specific environmental priors\cite{zhao2023common, long2022environment}. For localization, it enriches fingerprinting with diverse channel features, outperforming traditional signal-strength-based fingerprinting methods\cite{wu2023environment2}. For wireless sensing, a particular type of CKM called clutter angular map (CLAM) was proposed in \cite{xu2024channel}, which learns the clutter angular profiles to spatially nullify interference, enhancing the sensing performance particularly for low-speed targets in cluttered environments. 

To practically unlock the full potential of CKM for wireless systems exemplified above, one central issue lies in how to construct CKM efficiently.
There are two major categories for CKM construction, namely model-based and model-free approaches.
Model-based CKM construction utilizes well-established channel models and some on-site channel knowledge data to solve curve fitting problems in order to achieve low-complexity CKM construction method with higher generalization capability\cite{ckm}.
For example, the authors in \cite{xu2024much} offered a preliminary analysis of constructing channel gain map (CGM) by utilizing the classic path loss model based on a limited number of channel gain data. 
Meanwhile, a CKM construction approach based on the expectation maximization (EM) algorithm was put forward in \cite{li2022channel}, which makes use of measurement data together with expert knowledge from established statistical channel models. 
On the other hand, model-free CKM construction does not rely on channel models, but directly utilizing data to build the CKM. Traditional model-free CKM construction includes methods like inverse distance weighting (IDW)\cite{chiles2012geostatistics}, K-nearest neighbor (KNN)\cite{cover1967nearest} and Kriging\cite{dall2011channel}. Such construction methods try to predict channel knowledge at unknown location through interpolation with existing data. Recently, with the advancement of AI, several deep neural networks (DNNs) like RadioUNet\cite{levie2021radiounet} and RME-GAN\cite{zhang2023rme} were designed to construct CKMs.
In addition, the authors in \cite{wang2024deep} proposed a method using SRResNet to construct a complete CKM from sparse measurements, outperforming interpolation-based techniques without requiring additional inputs like physical environment maps. 
In \cite{fu2024generative}, a generative CKM construction method was proposed using a conditional diffusion model to reconstruct complete CKMs from partially observed data.
The authors in \cite{dai2024generating} proposed a CKM inference method using a UNet-based deep learning model to generate channel knowledge maps for new access points (APs) by leveraging CKMs and location information of existing APs.
Essentially, AI-based methods are also model-free approachs, but it usually requires a much greater amount and higher quality of numerical data than traditional interpolation methods like KNN. 
Therefore, there is no doubt that high quality data is essential for efficient CKM construction and utilization.

In terms of channel knowledge data acquisition, one may tend to utilize those channel models defined by 3GPP\cite{3gpp}. These models, while played crucial role for advancing wireless research and standardization, are usually unable to fully capture the complexities of actual propagation environment due to their statistical nature. Therefore, channel knowledge data directly generated by such models may not serve the purpose for AI-based CKM construction and utilization. Instead, it is imperative to develop a dataset that can offer data reflecting the actual propagation environment.
In fact, the development of high-quality datasets has been a cornerstone in advancing research across various domains, particularly in computer vision (CV). In natural image processing, the introduction of ImageNet\cite{deng2009imagenet} marked a transformative milestone by providing a large-scale, hierarchical image database with millions of high-quality, human-annotated images across thousands of categories. ImageNet's unprecedented scale and diversity make it a pivotal resource for tasks such as object recognition, classification, and localization, catalyzing significant advancements in AI-driven image understanding and deep learning techniques like convolutional neural networks (CNNs). Following ImageNet, several other influential datasets have been developed, each addressing specific challenges and expanding the scope of CV research. For instance, COCO \cite{lin2014microsoft} introduced a focus on object detection, segmentation, and captioning in complex scenes, emphasizing contextual understanding. Similarly, Pascal VOC\cite{everingham2010pascal} played a crucial role in early object detection and segmentation tasks, providing a benchmark for evaluating model performance in real-world scenarios. More recently, Open Images\cite{kuznetsova2020open} has further pushed the boundaries by offering a massive dataset with annotations for object detection, segmentation, and visual relationship detection.

In wireless community, recent efforts have also been devoted to dataset development to bridge AI and physical-layer wireless communication research. For instance, DeepMIMO\cite{alkhateeb2019deepmimo} provides synthetic mmWave channel data generated via ray-tracing. DeepSense 6G\cite{alkhateeb2023deepsense} integrates multi-modal sensing data with communication measurements, and RadioMapSeer\cite{yapar2022dataset} offers two-dimensional (2D) path loss maps for urban environments. These datasets, while valuable, usually only concern rather limited channel knowledge type, such as the path loss or channel gains. Besides, most of them do not provide the location-specific channel knowledge data, i.e., channel data item is only provided without tagging the absolute or relative locations of the mobile devices. Such limitations hinder their ability to fully capture the complex interplay between wireless channel knowledge and local wireless environment, or to enable the AI-based CKM construction for more diversified channel knowledge types beyond channel gains. 

To fill the above gap, inspired by the great success of ImageNet, in this paper, we propose CKMImageNet\textsuperscript{1} \footnotetext[1]{\url{https://github.com/Darwen9/CKMImagenet}}, a comprehensive data set designed to bridge AI and environment-aware wireless communications and sensing, by integrating location-specific channel knowledge data, high-fidelity environmental maps, and visual representations (e.g. grayscale images). The main contributions of this paper
are summarized as follows:

\begin{itemize}
    \item Unlike existing datasets like RadioMapSeer\cite{yapar2022dataset} that only focuses on isolated parameters (e.g., signal power or path loss), CKMImageNet provides multi-dimensional channel knowledge spanning channel gain, AoA, AoD, propagation delay, and other key parameters. This holistic representation enables full-stack analysis of wireless channel knowledge, supporting diverse AI-driven tasks from environment-aware network planning to proactive resource allocation.
    \item CKMImageNet introduces normalized, spatially consistent and location-specific grayscale images ($32\times32$, $64\times64$, $128\times128$ pixels) encoding multi-dimensional channel parameters. This innovation bridges numerical channel knowledge data with vision-based AI architectures, reducing computational complexity while preserving spatial correlations between channel characteristics and environmental features. Compared to prior works with limited visual formats, this approach ensures compatibility with standard computer vision pipelines and enhances interpretability.
    \item CKMImageNet pioneers the tight coupling of numerical channel data, binary obstacle matrices, and grayscale CKMs into a unified framework. By integrating explicit environmental context (e.g., building layouts, material properties), the dataset enables AI models to learn causal relationships between physical environments and propagation channels.
\end{itemize}

The organization for the rest of this paper is as follows. 
Section II reviews related wireless datasets and compare our proposed CKMImageNet dataset with them.
Section III presents the details of the CKMImageNet dataset.
In Section IV, we provide a detailed introduction to the method of constructing the dataset.
Section V describes some specific use case applications of the dataset. 
Finally, Section VI concludes this work with future plans.

\section{Related wireless datasets}
In this section, several related wireless datasets are introduced. Additionally, the proposed CKMImageNet dataset is compared with existing ones, and the differences are summarized in Table.~\ref{tab:comparison}.

\subsection{DeepMIMO}
The authors in \cite{alkhateeb2019deepmimo} introduced a dataset named DeepMIMO. It is a highly flexible and parameterized dataset designed for research in mmWave and multiple-input multiple-output (MIMO) systems. It uses ray-tracing simulations to generate channel data that captures the dependence on environmental geometry, materials, and user positions. Researchers can customize parameters such as antenna configurations, bandwidth, and the number of paths, making it a versatile tool for tasks like beam prediction and channel estimation. However, DeepMIMO primarily focuses on synthetic data for predefined scenarios without offering visual representations or physical environment maps.

\subsection{DeepSense 6G}
The DeepSense 6G dataset \cite{alkhateeb2023deepsense} provides real-world multi-modal data by combining mmWave measurements, LiDAR, RGB images, and global positioning system (GPS) coordinates. It supports applications that integrated sensing and communication (ISAC), such as vehicular communication. While its rich multi-modal nature is valuable for ISAC research, the data it provides for communication is very limited as it only contains signal power.

\subsection{RadioMapSeer}
The RadioMapSeer dataset \cite{yapar2022dataset} focuses on urban environments with transmitters located at ground level and realistic urban features, such as cars, to simulate traffic and its impact on radio propagation. It provides path loss radio maps and utilizes models like dominant path model (DPM) and intelligent ray-tracing (IRT) to predict radio coverage. While it is effective for studying path loss and understanding propagation characteristics, its focus on 2D path loss limits its applicability to more complex, multi-dimensional wireless modeling.

\subsection{3DRadioMap}
The 3DRadioMap dataset \cite{yapar2022dataset} extends to three-dimensional modeling by including rooftop transmitters and varying building heights, resulting in height-encoded radio maps. This dataset is designed for localization studies, utilizing both path loss and time-of-arrival (ToA) data to evaluate position estimation methods. The three-dimensional (3D) nature of the data allows for the exploration of complex phenomena like reflections and shadowing, making it suitable for detailed propagation and localization studies. However, it lacks the diversity of data formats or the integration of additional channel knowledge such as angle information and its application is limited. 

\subsection{$\textbf{M}^{\textbf{3}}\textbf{SC}$}
The authors in \cite{cheng2023m} proposed the $\mathrm{M}^{3}\mathrm{SC}$ dataset, which is a simulation based dataset designed for mixed multi-modal (MMM) sensing and communication integration. The dataset includes a rich collection of multi-modal sensing information such as RGB images, depth maps, LiDAR point clouds, and mmWave radar point clouds, along with communication data like channel impulse response (CIR) matrices. $\mathrm{M}^{3}\mathrm{SC}$ is generated using advanced simulation tools like AirSim for sensory data and Wireless InSite for communication data, ensuring high fidelity and customization. However, it lacks comprehensive channel knowledge types(e.g., AoA, AoD) and is limited to vehicular urban scenarios.

\vspace{1.0em}
Based on the above analysis, existing wireless datasets have played a significant role in specific scenarios or tasks. However, they share several common limitations that hinder their ability to fully capture the complexities of real-world wireless environments. They tend to concentrate on single type of channel knowledge like signal strength, rather than incorporating multi-dimensional channel knowledge. Also, these datasets usually lack visual data formats (e.g., grayscale images) that are important for AI models, especially computer vision algorithms, to learn spatial correlations and environment-channel knowledge interactions effectively. Moreover, most of them do not include environmental maps (e.g., building layouts, material properties), which makes them difficult to model the causal relationships between channel knowledge and the physical environment.

\begin{table*}[htbp]
    \centering
    \caption{Comparison of relevant wireless datasets and our proposed CKMImageNet.} 
    \label{tab:comparison}
    \vspace{10pt} 
    \resizebox{\textwidth}{!}{
    \begin{tabular}{lcccccc}
        \toprule
         & DeepMIMO ~\cite{alkhateeb2019deepmimo} & DeepSense 6G ~\cite{alkhateeb2023deepsense} & RadioMapSeer ~\cite{yapar2022dataset} & 3DRadioMap ~\cite{yapar2022dataset}& M$^3$SC ~\cite{cheng2023m} & Proposed CKMImageNet\\
        \midrule
        Data with location information  & $\times$ & $\times$ & \checkmark & \checkmark & $\times$ & \checkmark\\
        Signal power/channel gain  & \checkmark & \checkmark & \checkmark & \checkmark & \checkmark & \checkmark\\
        Angle information  & \checkmark & $\times$ & $\times$ & $\times$ & $\times$ & \checkmark\\
        Propagation delay  & \checkmark & $\times$ & $\times$ & $\times$ & $\times$ & \checkmark\\
        Visual CKM  & $\times$ & $\times$ & \checkmark & \checkmark & $\times$ & \checkmark\\
        Environment information integration  & $\times$ & \checkmark & \checkmark & \checkmark & \checkmark & \checkmark\\
        Customization for application requirement  & \checkmark & \checkmark & \checkmark & \checkmark & \checkmark & \checkmark\\
        \bottomrule
    \end{tabular}
    }
\end{table*}

\section{What is CKMImageNet}

\subsection{Content and Framework of CKMImageNet}

CKMImageNet is a dataset designed to advance AI-driven environment-aware wireless research by providing location-specific channel knowledge data. It is generated based on high-fidelity electromagnetic tray tracing computations instead of relying on statistical channel models such as Rician, Rayleigh fading model or geometric-based stochastic model (GBSM). CKMImageNet provides spatially consistent and location-specific channel knowledge with real-world environmental geometries, enabling AI models to learn propagation dynamics governed by physical laws rather than statistical approximations.

The core of CKMImageNet lies in high-fidelity generation framework. Different types of channel knowledge are generated using ray-tracing tools such as \textit{Wireless Insite}\textsuperscript{2} \footnotetext[2]{\url{https://www.remcom.com/wireless-insite-em-propagation-software}} and \textit{Sionna}\textsuperscript{3}\footnotetext[3]{\url{https://github.com/NVlabs/sionna}}. These tools model electromagnetic wave interactions, such as reflections, diffractions, and scattering, with real-world 3D geometries. The simulations account for material properties like conductivity and permittivity, as well as environmental features such as buildings and terrain, ensuring that the data aligns with real-world electromagnetic behavior. Each data instance is structured as a numerical data pairs $(\mathbf{q},\mathbf{z})$, where $\mathbf{q} \in \mathbb{R}^3$ represents the 3D spatial coordinates of a mobile receiver relative to a transmitter, capturing the absolute or relative positioning, $\mathbf{z} \in \mathbb{R}^d$ denotes multi-dimensional channel knowledge like channel gain, AoA, AoD and propagation delay, with $d$ representing the dimension of the channel knowledge. 

To enhance flexibility for AI-driven wireless communication tasks, channel knowledge values are normalized and spatially mapped into grayscale images with sizes $32\times32$, $64\times64$ and $128\times128$. This transformation to grayscale offers several important advantages. First, the standardized channel knowledge format provides a simplified yet effective data representation that reduces computational complexity compared to multi-channel alternatives, making it particularly suitable for resource-constrained AI applications. The grayscale format (0-255) also naturally aligns with normalized channel parameters, ensuring consistent feature scaling while avoiding potential biases from varying parameter ranges. Additionally, by eliminating color information, this approach focuses the model's attention on essential spatial patterns of channel characteristics while reducing noise interference. The grayscale format further ensures compatibility with standard vision-based AI architectures, allowing seamless integration with popular pre-trained models without requiring extensive modifications. Meanwhile, the provision of multiple image resolutions deliver important flexibility for diverse implementation scenarios. Bigger sizes (e.g., 128×128) preserve fine-grained spatial details necessary for precision-demanding applications and smaller sizes (e.g., 32×32) significantly reduce computational overhead. Furthermore, this scalable design ensures future adaptability to evolving AI models and hardware capabilities without necessitating fundamental changes to the data processing pipeline, making it a robust and forward-compatible solution for AI-driven wireless communication systems. 

As illustrated in Fig.~\ref{frame}, the CKMImageNet framework adopts a hierarchical, net-like structure organized into three core components: scenarios, base stations (BSs), and data. Users can progressively navigate this hierarchy to select datasets tailored to their research needs. The final retrievable data comprises the following elements:

\begin{figure*}[htbp] 
\centering 
\includegraphics[width=\textwidth]{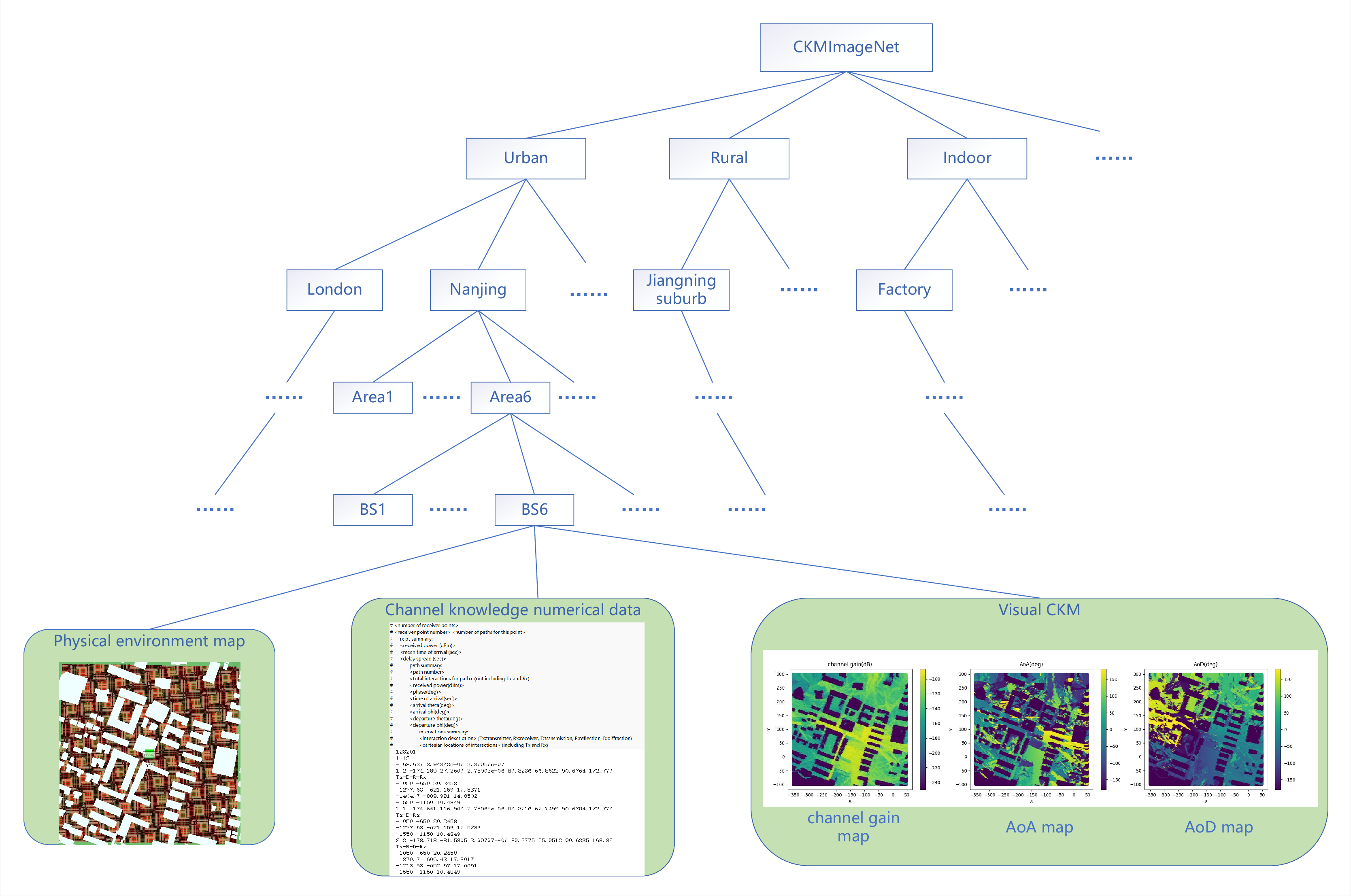} 
\caption{Framework of CKMImageNet.} 
\captionsetup{font=normalsize}
\label{frame} 
\end{figure*}

\subsubsection{Physical Environment Maps}The foundation of CKMImageNet lies in its high-fidelity physical environment maps, which digitize real-world spatial and structural features to reflect electromagnetic propagation conditions. These maps are derived from diverse global scenarios, including urban, rural, and indoor environments, and incorporate detailed 3D geometries of buildings, terrain profiles, and material properties (e.g., conductivity, permittivity). To date, the dataset includes over 10,000 environment maps, each spanning 128m × 128m, generated using tools like \textit{Blender}\textsuperscript{4} \footnotetext[4]{\url{https://www.blender.org/}} and \textit{OpenStreetMap}\textsuperscript{5} \footnotetext[5]{\url{https://www.openstreetmap.org/}}. These maps are stored as binary matrices $\mathbf{M} \in \{0,1\}^{H \times W}$, where 1 indicates obstacles (e.g., buildings, vehicles), and $H$ and $W$ indicates the approximate scale of the physical map.

\subsubsection{Channel Knowledge Numerical Data}
CKMImageNet houses multi-dimensional channel data knowledge numerical including channel gain, AoA, AoD and propagation delay. And the numerical data is stored in .p2m files, which contain channel knowledge at various location points within the simulation area. 

\subsubsection{Visual CKM}
To bridge AI and environment-aware wireless communication research, CKMImageNet transforms channel parameters into grayscale images for intuitive analysis and AI training. Each image spatially maps normalized channel data (e.g., channel gain, angles) to pixel intensities (0–255), with obstacles fixed at 0 (black).

The framework's components are tightly integrated to form a cohesive system where environment maps are utilized as the initial input for ray-tracing simulations. These simulations are instrumental in generating channel knowledge data, which is then spatially mapped and normalized to produce grayscale images. These images serve a crucial role in visually connecting environmental features, such as building layouts, with channel characteristics, including signal attenuation zones. This visual linkage facilitates an AI-driven approach to environment-aware communication and sensing tasks ranging from network planning such as communication BS site selection and sensing anchor node placement, to pro-active resource allocation such as beam alignment, power allocation, interference avoidance, clutter rejection, and robot trajectory planning. Furthermore, the hierarchical structure of the framework ensures seamless access to multi-modal data, which is essential for empowering researchers to effectively design, analyze, and optimize environment-aware 6G systems. This integrated approach not only streamlines the process but also enhances the overall efficiency and effectiveness of the system.

\subsection{Key Features of CKMImageNet}

\textbf{Data with spatial consistency:} CKMImageNet ensures high fidelity and spatial consistency by leveraging physics-based ray-tracing and real-world measurements. Instead of using data based on statistical models (e.g., Rician, Rayleigh) or simplified geometric models, CKMImageNet integrates high-resolution environmental maps (e.g., building layouts and material properties) to accurately capture spatial correlations in signal propagation. For instance, phenomena such as signal attenuation and diffraction are modeled in alignment with the actual environmental topology, avoiding the oversimplifications of traditional approaches. This makes CKMImageNet a robust and realistic dataset for capturing the complexities of real-world wireless communication scenarios.

\textbf{Location-Channel Knowledge Pairing:} Each data entry $(\mathbf{q},\mathbf{z})$ explicitly links channel parameters to geographic or relative coordinates, enabling precise modeling of spatial variations in signal propagation. This correlation is critical for constructing environment-aware CKMs.

\textbf{Multi-Dimensional Channel Knowledge:} CKMImageNet includes multi-dimensional channel parameters, which can support a full-stack analysis ranging from physical propagation to network optimization. In detail, the dataset contains $\mathbf{z} = \begin{bmatrix} g, \theta_{\text{AoA}}, \theta_{\text{AoD}}, \tau, ... \end{bmatrix}^T \in \mathbb{R}^d$, where $g$ represents channel gain, $\theta_{\text{AoA}}$ represents  angle of arrival, $\theta_{\text{AoD}}$ represents angle of departure, $\tau$ represents propagation delay. And these multi-dimensional channel knowledge represent the information of each path, with a total of 25 paths containing such information. This multi-dimensional approach allows for a comprehensive analysis of the wireless channel, supporting various environment-aware wireless tasks like network planning and resource allocation.

\textbf{Visual CKM Representations:} CKMImageNet offers a range of visual
CKM, which provides an intuitive understanding of channel knowledge, making it easier for users to interpret and utilize the data. In addition, visualizing these numerical datasets can better integrate with AI image processing as will be discussed in Section~\ref{sec:case}.

\textbf{Environment-Data Fusion:} Physical environment maps (e.g., building layouts) are integrated to enable AI models to learn context-dependent propagation patterns. 

\textbf{Multi-Scenario support:} CKMImageNet is designed to
support a wide variety of scenarios, encompassing urban, rural and indoor environments. This multi-scenario
support allows researchers and engineers to analyze and
optimize wireless networks across different contexts. The
ability to simulate and study various environments makes
the CKMImageNet a versatile tool, useful for a broad
range of applications and research projects.

\section {Construction Method of CKMImageNet}

In this section, the methodology for constructing the CKMImageNet dataset is introduced, which integrates three primary components: simulation data, real-world measurement data, and a collaborative data-sharing platform. The simulation data is generated using advanced ray-tracing tools, including Wireless Insite and the open-source Sionna framework, to model high-fidelity electromagnetic wave propagation across diverse environments. These tools capture detailed interactions such as reflections, diffractions, and scattering, ensuring spatially consistent channel knowledge aligned with real-world physical laws. Complementing simulations, measurement campaigns collect location-specific channel parameters (e.g., signal strength, AoA and AoD) to incorporate empirical data, enhancing the dataset’s realism and diversity. Finally, the CKMImageNet platform facilitates collaborative data sharing, enabling researchers worldwide to contribute and utilize standardized datasets. This tripartite approach—combining physics-based simulations, empirical measurements, and community-driven expansion—ensures the creation of a comprehensive, scalable, and environment-aware resource for advancing AI-driven 6G communication research.

\subsection{Simulation Data}
\begin{figure*}[htbp] 
\centering 
\includegraphics[width=0.8\textwidth]{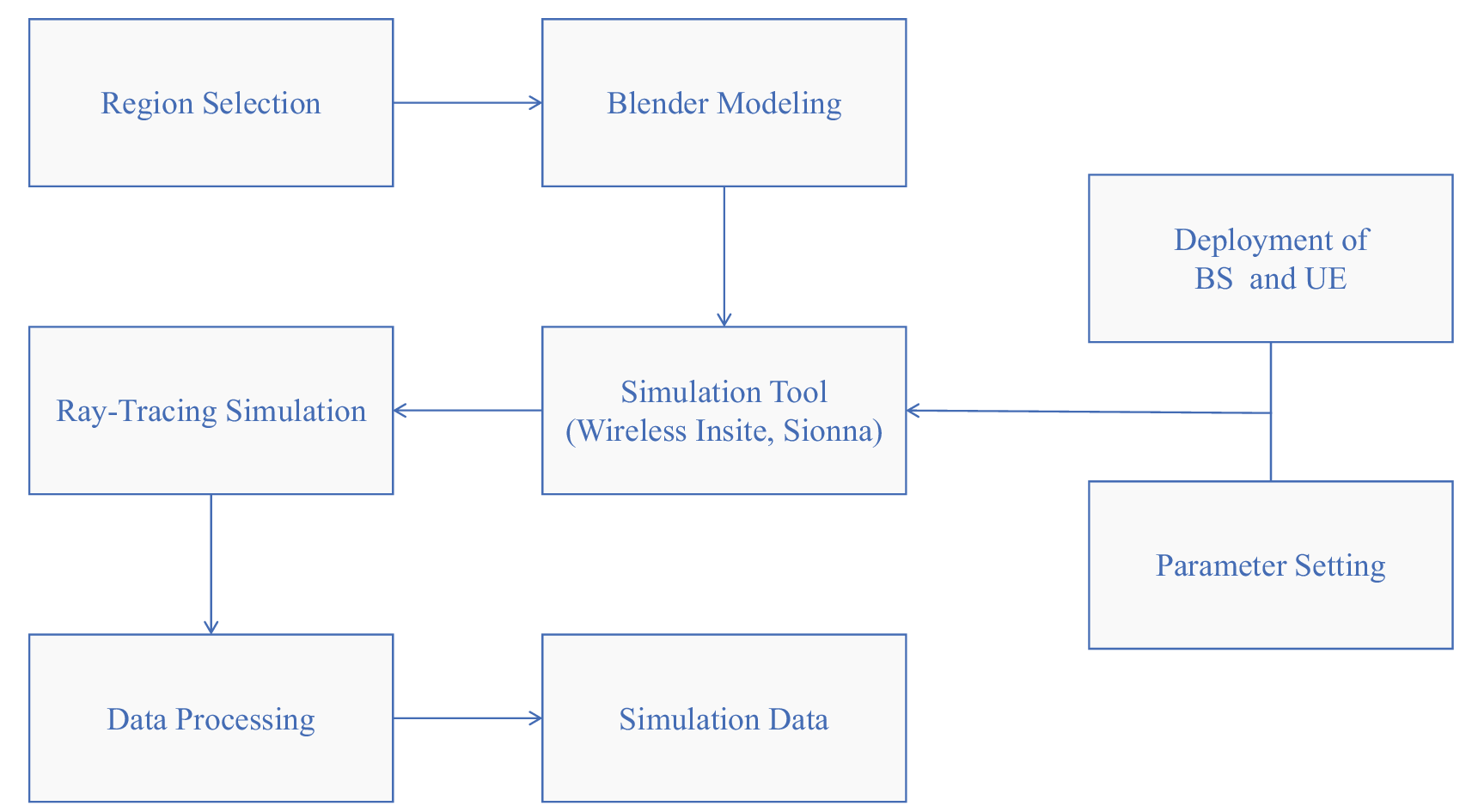} 
\caption{The main procedures for generating channel knowledge data for CKMImageNet.} 
\label{process} 
\end{figure*}

In this section, we will describe in detail the different elements of the CKMImageNet dataset construction process, and the process of simulation construction is shown in Fig.~\ref{process}. When we initially construct the CKMImageNet dataset, we directly import the scenario models like Beijing, Boston and London provided by the official website of Wireless Insite into the software. However, the scenarios are limited, which fail to generate enough data. Here, we introduce a method to produce more data to augment the CKMImagenet dataset. Following is an example of constructing CKMImageNet of the region of a neighbourhood in Nanjing, China.   

\subsubsection{Select Region}
\begin{figure}[htbp] 
\centering 
\includegraphics[width=0.5\textwidth]{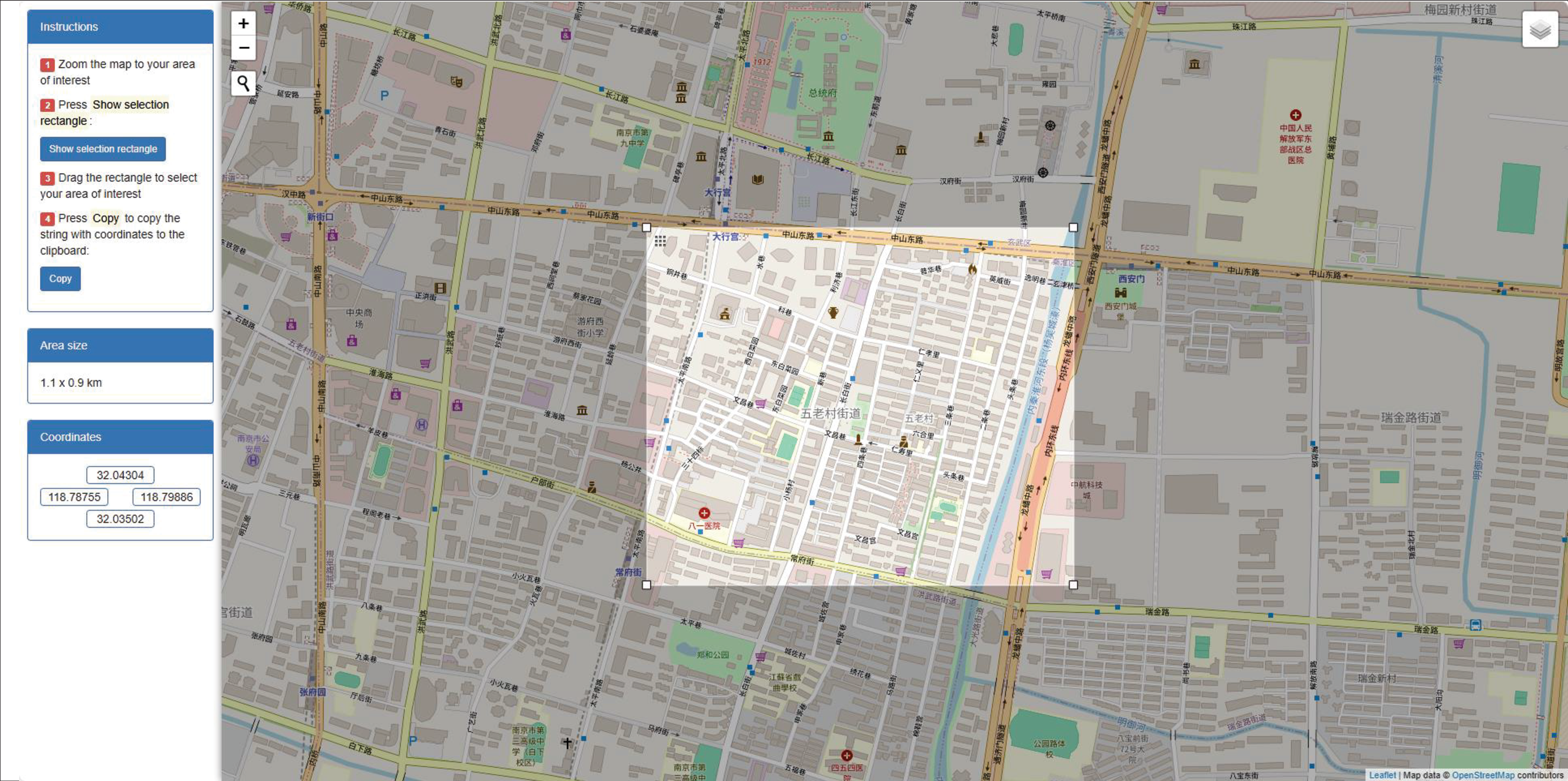} 
\caption{An illustration of selecting region in OSM.} 
\label{osm} 
\end{figure}

\begin{figure}[htbp] 
\centering 
\includegraphics[width=0.5\textwidth]{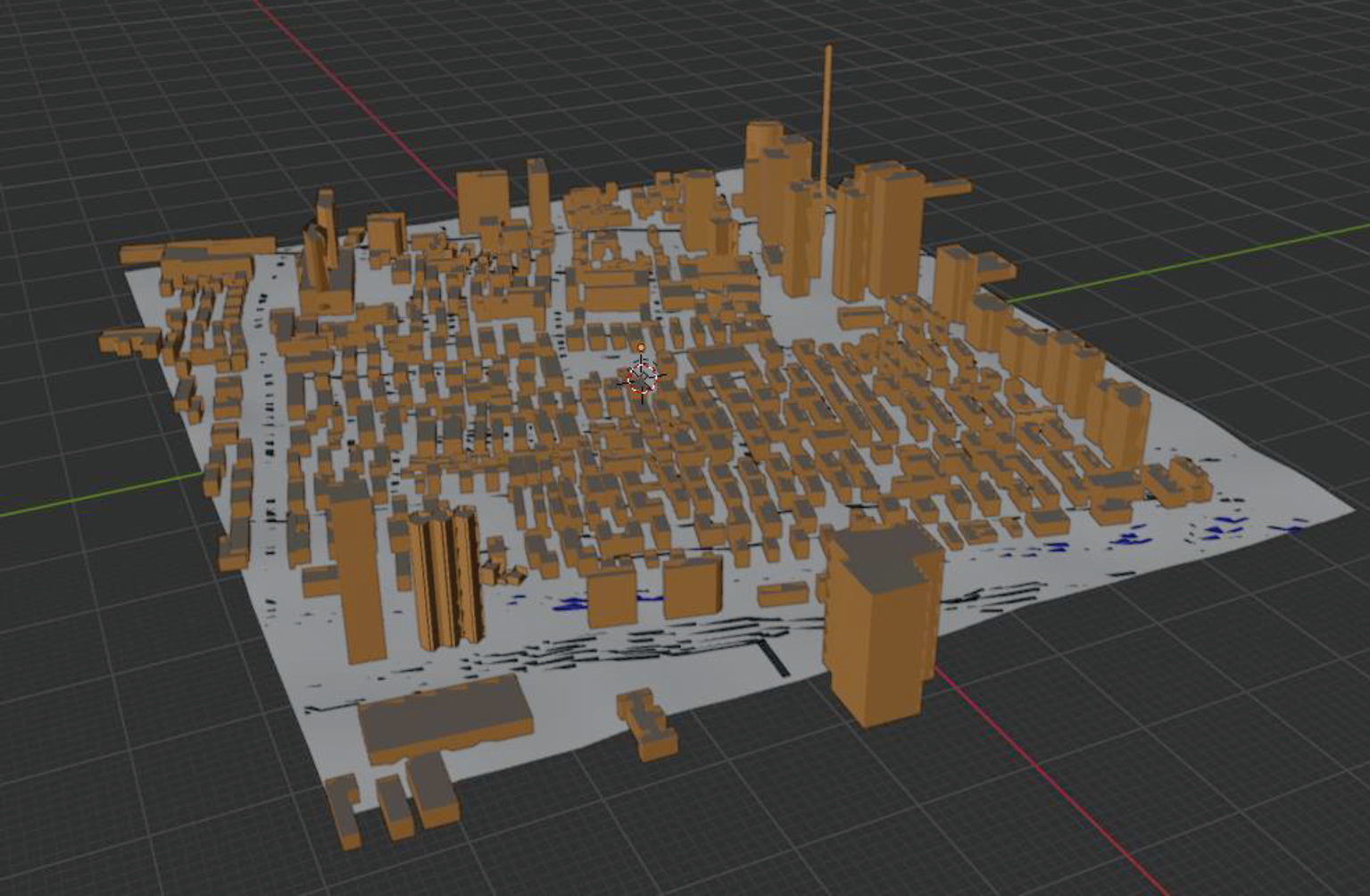} 
\caption{Blender modeling of a region.} 
\label{blender} 
\end{figure}

In order to construct CKMImageNet, we need a 3D environment map and import it into ray tracing software like Wireless Insite, and then the following steps can be performed. To complete this step, the professional modeling software Blender is of great help. First, we need to install an add-on named blender-osm from which the extent of OpenStreetMap can be obtained. Once the add-on is installed, we can select a region from OpenStreetMap, which is a map of the world and contains information about buildings and roads in most parts of the world. With the assistance of the tool, it is accessible for us to select region of interest around the world. As shown in Fig.~\ref{osm}, a rectangular area of Nanjing is selected, then the environment map is imported into Blender, in which the 3D architecture of the area can be modeled, as shown in Fig.~\ref{blender}. Finally, export the 3D environment model as Collada.dae file, and then it can be opened in Wireless Insite for the subsequent operation.

\subsubsection{Wireless Insite Parameter Setting and Ray-tracing Simulation}
\begin{figure}[htbp] 
\centering 
\includegraphics[width=0.5\textwidth]{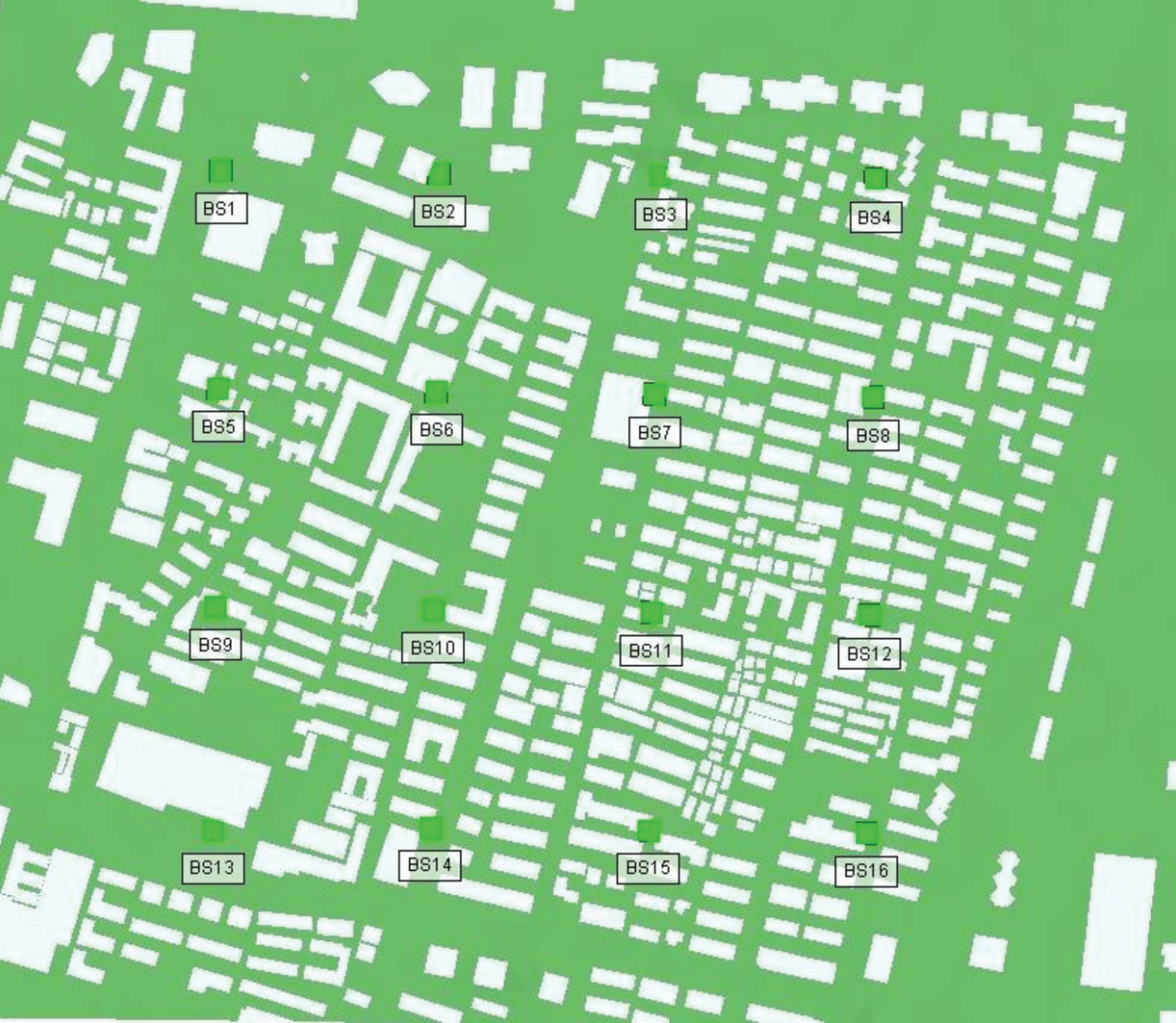} 
\caption{Wireless Insite region deployment scenario.} 
\label{BS} 
\end{figure}

Once the 3D environment model is imported into Wireless Insite, the environment map of the region can be seen as shown in Fig.~\ref{BS}  and the next step is to set up the BS and user equipment (UE) within the environment and configure the parameters of the simulation scenario.

The maps imported into the simulation software are typically large in scale. To optimize simulation efficiency, we divide them into smaller study areas. This approach accelerates the process, as the software performs ray-tracing from each transmitter to every deployed receiver within a study area. Generally, we place a $4\times4$ grid of BSs within a $300 \times300 m^2$ rectangular area, with adjacent BSs spaced $100m$ apart. To ensure comprehensive coverage, we extend this area by $200m$ on each side, forming a larger $700 \times700 m^2$ rectangle. Receivers are then deployed every $2m$ within this extended area, resulting in a grid of $351\times351=123,201$ receivers. This setup ensures that every transmitter has receivers within a $400 \times400 m^2$ coverage area. Each study area thus produces up to $16\times123,201$ data points, excluding locations where receivers are blocked by buildings. Completing ray tracing for such a study area typically requires an entire day to complete.

We select a signal frequency of 28GHz, and each BS and UE have only a single half-wave dipole with the axis of the dipole antenna in the z-direction.In addition, the maximum number of multipath components is set to 25, with a maximum of 6 reflections and 1 diffraction to simulate real-world scenarios. Detailed simulation parameters are provided in the table.~\ref{ckm_parameter}

\begin{table}[!t]
 \renewcommand{\arraystretch}{1.3}
 \caption{CKMImageNet Simulation Parameters}
 \centering
 \label{ckm_parameter}
 \begin{tabular}{c c}
 \hline\hline \\[-4mm]
\multicolumn{1}{c}{Parameter} & \multicolumn{1}{c}{value} \\[0.5ex] \hline
 Waveform  & Sinusoid \\
 Antenna & Isotropic \\
 Number of reflections  & $6$ \\
 Number of diffractions & $1$ \\
 X-axis minimum interval units  & $\Delta x = 2 $m  \\
 Y-axis minimum interval units  & $\Delta y = 2 $m  \\
\hline\hline
\end{tabular}
\end{table}

Once the BSs and UEs are deployed and the parameter configuration is completed, ray-tracing simulation is conducted.
Ray tracing is a technique to simulate environment-specific and physically accurate channel realizations for a given scene and user position.
As discussed in Section II, CKMImageNet utilizes advanced ray-tracing techniques to simulate electromagnetic wave propagation with high fidelity by using the commercial ray-tracing software Wireless InSite. Ray-tracing models the interactions of waves with the environment, including reflections and diffractions. This high level of fidelity ensures that the generated channel data is reliable, providing a solid foundation for advanced network design and analysis. 

\subsubsection{Grayscale images}

Grayscale images play a pivotal role in AI training, particularly in the realms of computer vision and image processing. By reducing color images to shades of gray, these images simplify the complexity of the data, allowing AI models to focus on the structural and textural aspects of visual data without the distraction of color. This simplification is crucial for training models to recognize patterns, edges, and shapes, which are fundamental to tasks such as object detection, facial recognition, and scene understanding. Moreover, working with grayscale images can significantly reduce the computational load, enabling faster training and more efficient use of resources. As a result, grayscale images are often used as a starting point in AI training pipelines, providing a solid foundation for more complex color image analysis. Since one of our main tasks is to leverage AI to better construct CKM, grayscale images of channel information are indispensable. The main generation steps are discussed as follows.

\textbf{Normalization}
Since each pixel value in a grayscale image ranges from 0 to 255, with 0 representing black and 255 representing white, we need to normalize the channel knowledge values to this range before proceeding to the subsequent operations. 
For channel gain grayscale images, since we have set the reception threshold at -250 dB and the maximum channel gain does not exceed -50 dB, we linearly normalize this range to 0-255. 
And the process can be expressed as: 
\begin{equation}
    \bar{x} = \frac{255(x + 250)}{200},
    \label{eq:normalization}
\end{equation}
where $x$ and $\bar{x}$ denote the real value of channel gain and the normalized value of channel gain, respectively. 

Additionally, for the buildings in the scenarios, since users cannot be placed at these locations and channel information cannot be obtained there, we consider the channel gain at these places to be the minimum of -250 dB. One point that may be questioned is whether this might be confused with the locations corresponding to the minimum channel gain -250 dB. Although the signal will attenuate significantly beyond the area we have selected, within the area it is very difficult for the signal to attenuate to -250 dB, so the probability of confusion is extremely small. Moreover, data indicates that within the range of most BSs, the minimum signal strength is generally above -200 dB. Therefore, such a setting is reasonable. 
For angle grayscale images, the range of angles is from \SI{-180}{\degree} to \SI{180}{\degree}. It is worth noting that for the buildings in the scenarios, in order to distinguish them from the locations that can receive signals and represent them in black in the grayscale image. We set their value to \SI{-200}{\degree} and linearly normalize the angle ranging from \SI{-200}{\degree} to \SI{180}{\degree} to 0-255, the process is similar to Formula~\ref{eq:normalization}.

\begin{figure*}[!h] 
\centering 
\includegraphics[width=\textwidth]{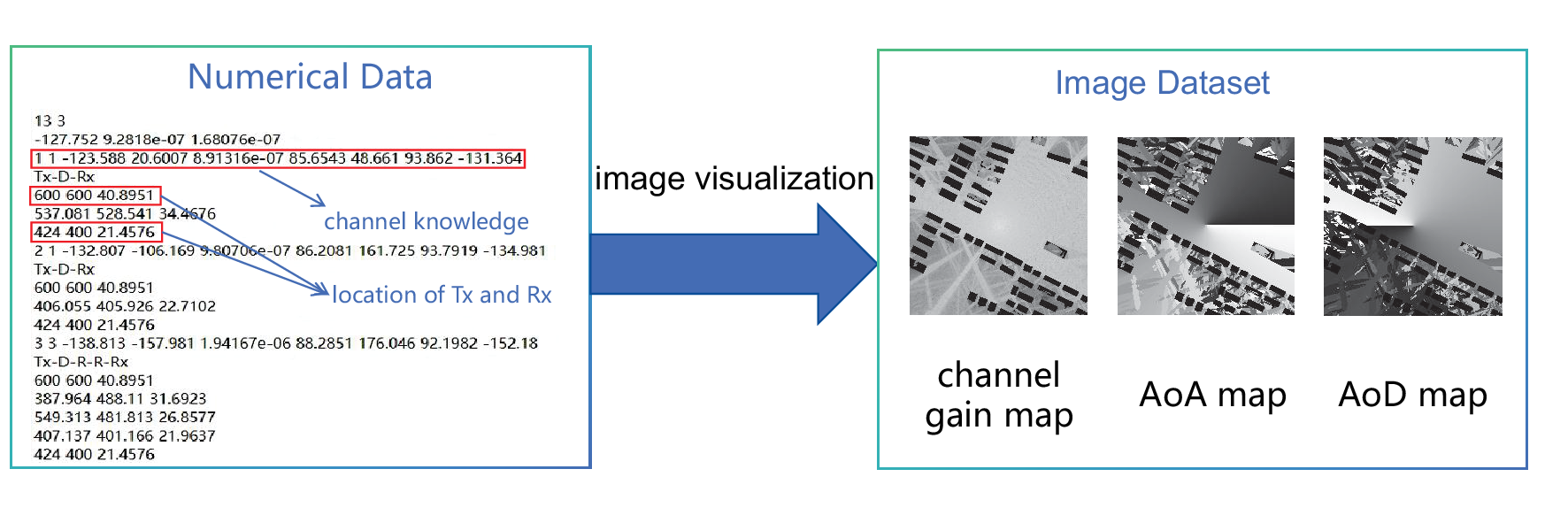} 
\caption{Visualization conversion from numerical data to grayscale images.} 
\label{visual} 
\end{figure*}

\textbf{Visualization}
As shown in Fig.~\ref{visual}, after obtaining the original numerical data, we process it by filtering out the values we need (such as the position information of transmitter (Tx) and receiver (Rx), as well as various channel knowledge). Then, after normalizing the data, we visualize it as grayscale images using the position information $\mathbf{q}$ and the corresponding $\mathbf{z}$.

\begin{figure}[htbp] 
\centering 
\includegraphics[width=0.5\textwidth]{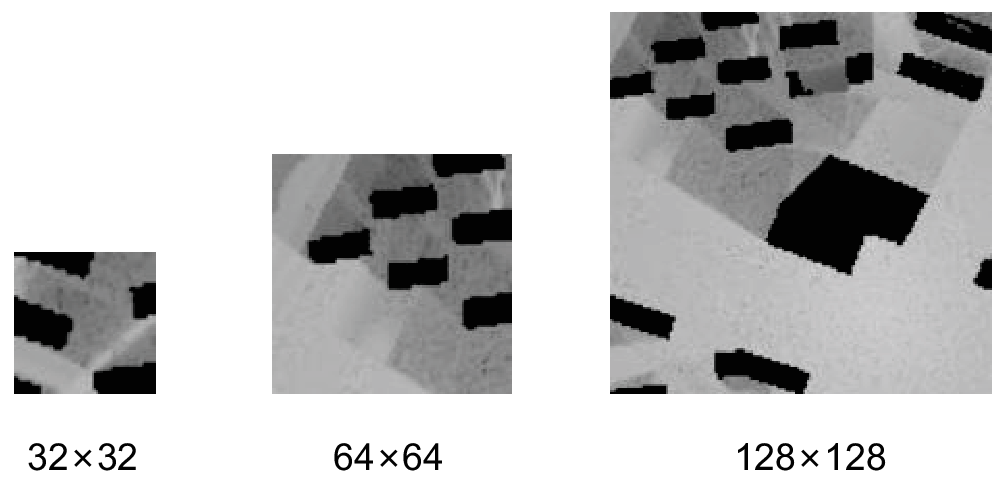} 
\caption{Different sizes of images.} 
\label{size} 
\end{figure}

\textbf{Size design}
We set the image size to $32\times32$, $64\times64$, $128\times128$ pixels as shown in Fig.~\ref{size}, with different sizes to meet different training needs. The reason for choosing image sizes that are powers of 2 is to enhance computational efficiency, storage efficiency, and compatibility with hardware and algorithm implementation, which are common practices in the field of AI image processing.  

\begin{figure}[htbp] 
\centering 
\includegraphics[width=0.5\textwidth]{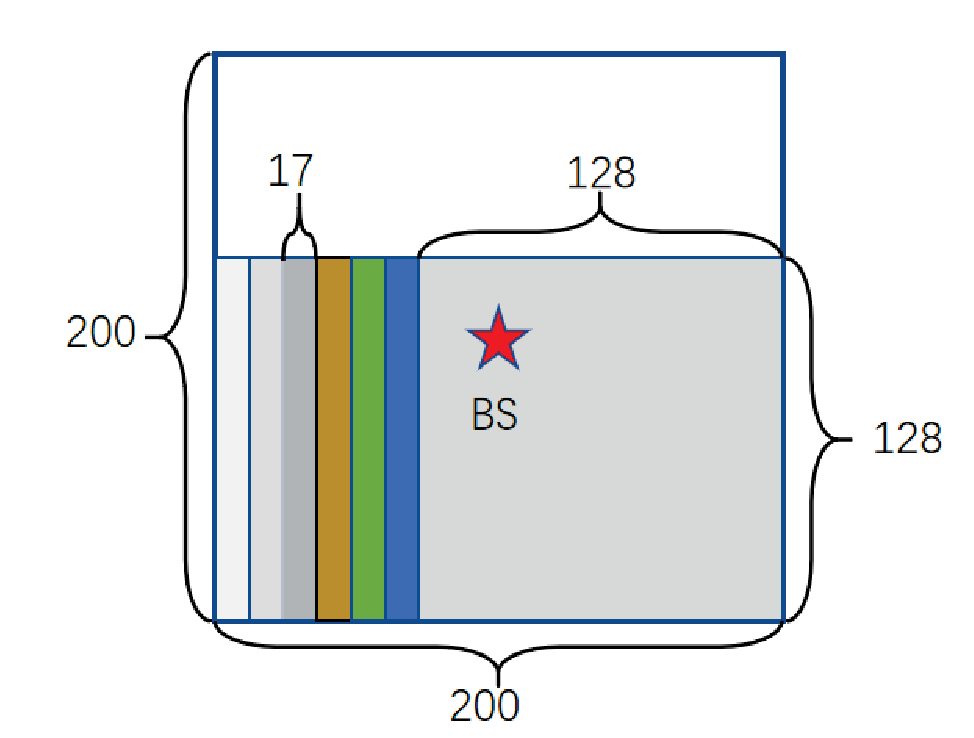} 
\caption{Demonstration of image segmentation.} 
\label{segment} 
\end{figure}

As for how to obtain these specific resolutions of images, the method we adopt is to segment from the complete images with the resolution of $200\times200$. For example, for images with the resolution of $128\times128$, the endpoints of two adjacent images are separated by 17 pixels, we can obtain 49 images with the resolution of 128 for the complete image, as shown in Fig.~\ref{segment}. Another reason for splitting images is that it can expand the dataset. If there are only complete images, the amount of data is too small to accomplish complex training tasks. 

\subsection{Sionna: An Open-Source Ray-Tracing Alternative}

 While commercial ray-tracing software like Wireless Insite offers robust capabilities, Sionna emerges as a powerful open-source alternative for constructing the CKMImageNet dataset. Developed as a graphics processing unit (GPU)-accelerated library specifically for next-generation physical layer research, Sionna is built on TensorFlow, enabling seamless integration of machine learning models and differentiable programming. This unique combination makes Sionna particularly valuable for AI-driven wireless communication research~\cite{hoydis2023sionna}.

\subsubsection{Coverage Map Generation} 
\begin{figure}[htbp] 
\centering 
\includegraphics[width=0.5\textwidth]{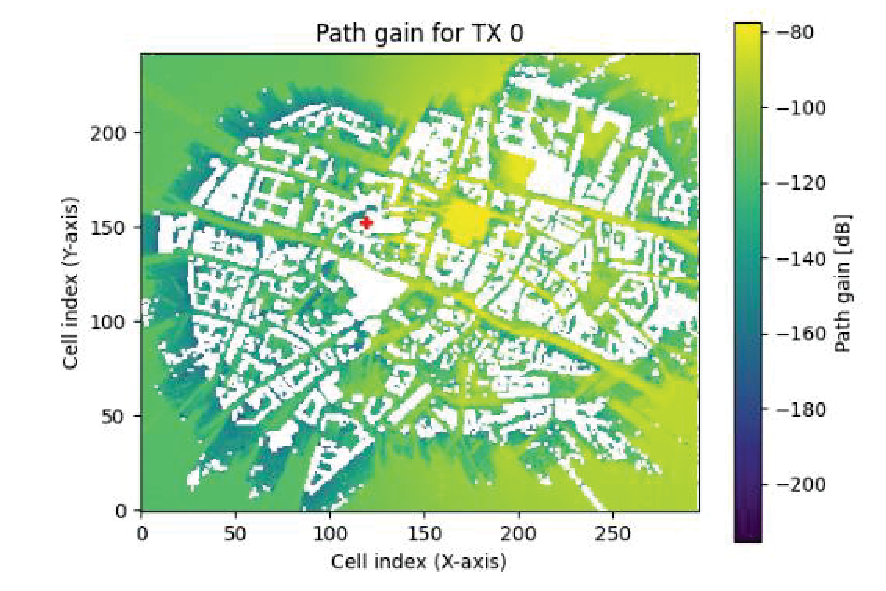} 
\caption{Covermap of Sionna} 
\label{covermap} 
\end{figure}

Sionna's coverage map functionality is essential for understanding radio wave propagation within a given environment. Users can define or load a scene, including physical layouts, material properties, and positions of transmitters and receivers. Sionna then computes propagation paths, tracing rays that represent radio waves and simulating their interactions with the environment, including reflections, diffractions, and scattering. The resulting coverage map visually represents signal strength across the environment, providing a comprehensive view of radio coverage, as shown in Fig.~\ref{covermap}.

\subsubsection{Point-to-Point Ray-Tracing} For each transmitter-receiver pair, Sionna performs detailed ray-tracing to simulate potential signal paths. This includes direct line-of-sight (LoS) paths and multi-path components resulting from environmental interactions. From these propagation paths, Sionna computes CIR, a crucial parameter in wireless communication systems that describes how the signal is affected by the environment, capturing the amplitude and delay of each path.

\vspace{0.5cm}

Sionna's integration with TensorFlow allows for the development of sophisticated AI models that can learn from and optimize wireless communication systems. The differentiability of Sionna enables gradient-based optimization of system parameters, such as antenna configurations and transmitter orientations, to maximize coverage or minimize interference. This feature is particularly valuable for training AI models to recognize patterns, predict channel behavior, and optimize network performance in real-world scenarios. Additionally, Sionna's open-source nature allows researchers to customize and extend its capabilities to suit specific research needs. This flexibility is crucial for developing innovative solutions in wireless communication and AI integration. Researchers can modify the ray-tracing algorithms, incorporate new environmental models, and integrate additional data sources to enhance the accuracy and applicability of their simulations.

In the context of CKMImageNet construction, Sionna's ability to generate detailed coverage maps and compute accurate CIRs is invaluable. These features provide the necessary data to create high-fidelity channel maps and visual representations, which are crucial for training AI models. Sionna's integration with AI frameworks also facilitates the development of advanced algorithms for channel estimation, beamforming, and interference management. Sionna represents a significant advancement in open-source tools for wireless communication research. Its combination of ray-tracing capabilities, AI integration, and flexibility makes it an essential tool for constructing comprehensive datasets like CKMImageNet. By leveraging Sionna, researchers can develop more accurate and realistic channel models, ultimately advancing the field of environment-aware 6G communication systems.

\subsection{Measurement Data}
Another important source of data comes from real-world measurement campaigns. During these measurements, we need to record key channel information such as signal strength, time of delay, AoA, and AoD and record the location information at different positions within the coverage area of a BS. Specifically, for each location, we capture its relative position coordinates with respect to the BS and measure the corresponding channel parameters. This raw data is then processed and stored in a predefined format, making it suitable for further analysis and integration into larger datasets. By doing so, we can expand this measured data into the CKMImageNet dataset, enhancing its comprehensiveness and diversity. 
Although we have not yet collected actual measurement data at this stage, measurement data can
not only help in enriching the dataset but also ensure that the channel characteristics captured from real-world environments are incorporated into the dataset, providing more accurate and realistic channel models for communication and AI research. This approach aligns with the growing need for diverse and high-quality data sources to improve AI models and algorithms in the communication domain, as highlighted in recent studies on data-driven wireless communication models. 

\subsection{Comparsion of data from different sources}
Sionna and commercial ray-tracing tools like Wireless Insite offer distinct advantages and limitations in CKMImageNet construction. Sionna, as an open-source, GPU-accelerated library built on TensorFlow, excels in seamless integration with AI frameworks, enabling differentiable ray tracing and gradient-based optimization for tasks like beamforming and network planning. Its open-source nature allows customization and extension, making it ideal for AI-driven research. However, Sionna’s pre-packaged functionalities limit flexibility: its coverage map generation only outputs signal strength, omitting multi-dimensional channel knowledge like AoA/AoD or propagation delay. For point-to-point ray-tracing, althought it can get multi-dimensional channel knowledge, it is less convenient for large-scale deployments of multiple Tx/Rx nodes compared to Wireless Insite. Sionna requires manual configuration via Python scripts, whereas Wireless Insite offers a more user-friendly graphical user interface (GUI) and automated batch processing for efficient network simulations. Additionally, Wireless Insite provides better operational ease for complex scenarios. Moreover, Wireless Insite provides high-fidelity, physics-based modeling of complex electromagnetic interactions (e.g., reflections, diffractions) across large-scale environments, supporting multi-BS deployments and generating rich channel knowledge (e.g., delay profiles, angles). Yet, its closed-source architecture limits customization, and the reliance on commercial licenses increases costs and accessibility barriers. 

As for measurement data, it offers unmatched realism by capturing ground-truth channel behavior influenced by dynamic environmental factors (e.g., moving objects, weather). This data is critical for validating simulation accuracy and training robust AI models. However, measurements are costly, time-intensive, and spatially limited, making large-scale coverage impractical. Sensor noise and calibration errors further degrade data quality. Conversely, ray-tracing data provides scalable, controlled simulations of diverse environments (urban, rural, indoor) with multi-dimensional channel parameters (e.g., AoA, delay) at minimal cost. By leveraging high-fidelity environmental maps, ray tracing ensures spatial consistency and supports scenario customization. However, its accuracy hinges on the precision of input models (e.g., material properties, geometry), and approximations in ray-tracing algorithms may introduce deviations from real-world propagation. While measurements anchor realism, ray tracing enables comprehensive, cost-effective dataset expansion—making their integration essential for balanced CKMImageNet construction.

\subsection{Data Sharing Platform}
Due to the limitations in the speed of generating simulation data and the high cost of acquiring real-world measurement data, we are going to develop a platform where we share our data generated through ray-tracing simulations and real-world measurements, along with algorithms for utilizing the CKMImageNet dataset. This platform is designed to better serve researchers interested in combining communication and AI, providing them with resources and tools to enhance their research. Given the limited capacity of individual researchers and the relatively small scale of our dataset compared to traditional computer science datasets, we encourage anyone with location-tagged data to contribute to CKMImageNet. By incorporating these data into the platform, we can process it and enrich the dataset, making it more comprehensive and valuable for future research in communication and AI integration. 

To ensure the quality and consistency of the data shared on the platform, we have established specific guidelines for data submission. These guidelines are designed to maintain the integrity of the CKMImageNet dataset and ensure that all contributions align with its core objectives.

\subsubsection{Data Format and Structure} For potential stakeholders who are willing to share data to the platform, we provide the format and structure of numerical data or grayscale images. All numerical data should be in the form of data pairs $(\mathbf{q},\mathbf{z})$ to ensure that the data is location-specific and can be easily integrated into the existing dataset. And images should be grayscale representations of channel knowledge, derived from the $(\mathbf{q},\mathbf{z})$ pairs to reflect the relative position relationships and channel characteristics.  

\subsubsection{Data Quality and Source} Data obtained from real-world measurement campaigns should be accurately recorded and processed. Key channel knowledge such as signal strength, propagation delay, AoA, and AoD should be measured at different positions within the coverage area of a BS. The raw data should be processed to remove noise and ensure accuracy before submission. And simulation data should be generated using advanced ray-tracing techniques that adhere to Maxwell's equations, ensuring high fidelity and environmental accuracy. The simulations should model real-world scenarios, including reflections, diffractions, and scattering, to provide a realistic representation of channel behavior. The use of commercial ray-tracing software like Wireless InSite is recommended to maintain consistency with the existing dataset.

\subsubsection{Environmental Context} While environmental context is highly valuable for enhancing the dataset's utility, it is not a mandatory requirement for data submission. However, if contributors choose to provide environmental context, it can significantly improve the dataset's ability to support environment-aware AI models and applications. If provided, environmental maps should detail the physical layout, material properties, and terrain features of the environment. And these maps should be stored as binary matrices $\mathbf{M} \in \{0, 1\}^{H \times W}$ where 1 indicates obstacles (e.g., buildings, walls) and 0 indicates free space. This information is crucial for enabling AI models to learn environment-aware propagation patterns.

\subsubsection{Data Sharing and Collaboration} The platform will provide open access to the shared data, allowing researchers to download and use the dataset for their studies. Contributors will be acknowledged for their contributions, and their data will be cited in any publications or research outputs derived from the dataset. In addition, the platform will facilitate collaborative research by connecting researchers with complementary expertise. For example, researchers with access to unique measurement environments can collaborate with AI experts to develop new algorithms for channel estimation, beamforming, or interference management.

By adhering to these guidelines, contributors can ensure that their data aligns with the core objectives of the CKMImageNet dataset, which is to bridge AI and environment-aware wireless communications and sensing by providing high-quality, and location-specific channel knowledge. This collaborative approach will not only enrich the dataset but also accelerate advancements in 6G communication systems, enabling more efficient and reliable network performance in real-world scenarios.

\section{Case Studies on How to Utilize CKMImageNet}
 
\label{sec:case}  

This section presents three case studies to illustrate how to utilize the developed CKMImageNet dataset, demonstrating its versatility in addressing critical challenges for environment-aware 6G communication and sensing systems. The case studies highlight how AI-driven approaches, powered by CKMImageNet’s multi-dimensional channel knowledge and visual representations, enable advanced solutions in diverse scenarios. Case 1 focuses on AI-based CKM construction, tackling challenges such as denoising, inpainting, and super-resolution of CKMs using generative models. Case 2 explores CKM-enhanced channel estimation, leveraging prior channel distributions derived from the dataset to improve accuracy in ultra-massive MIMO systems. Case 3 illustrates the integration of CKM with ISAC, where environmental context and channel features enhance sensing precision in complex urban environments. Together, these cases underscore CKMImageNet’s role as a foundational tool for advancing AI-driven innovations in environment-aware wireless networks.

\subsection{Case 1: AI-based CKM Construction}

\begin{figure*}[htbp] 
\centering 
\includegraphics[width=0.8\textwidth]{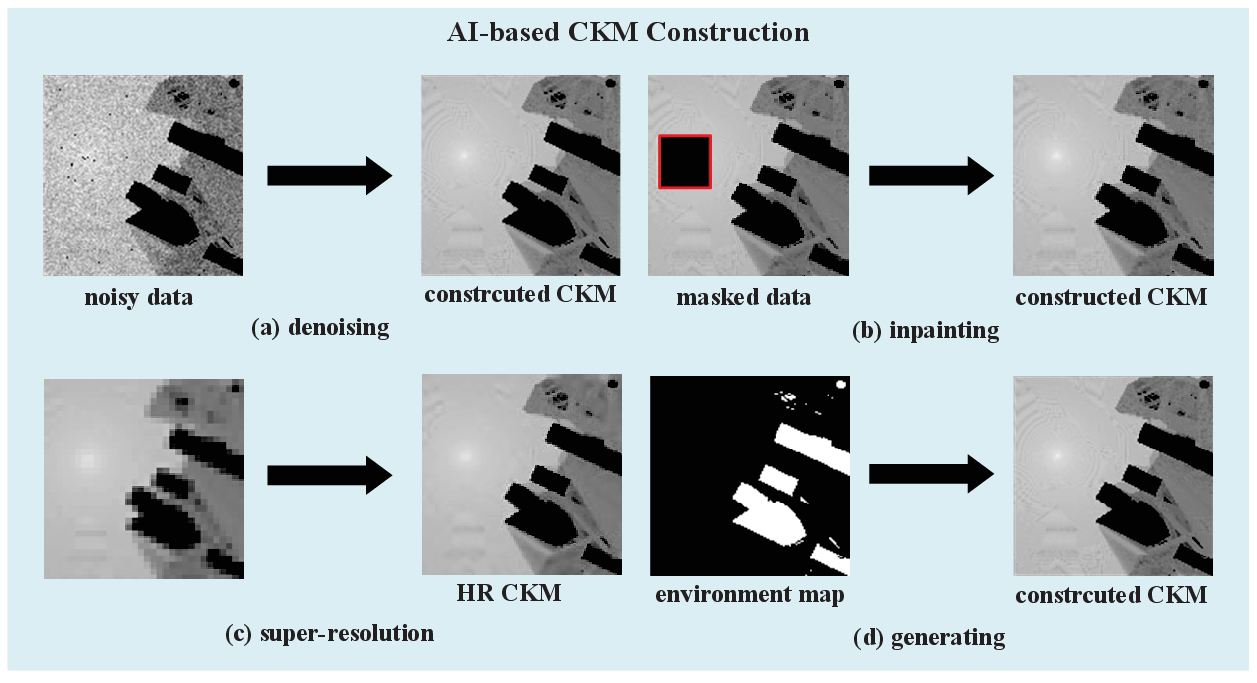} 
\caption{AI-based CKM Construction\cite{fushen}} 
\label{construct} 
\end{figure*}

The CKMImageNet dataset plays a pivotal role in addressing critical challenges in CKM construction,  especially when dealing with noisy, incomplete, or low-resolution data as shown in Fig.~\ref{construct}. It enables advanced AI-driven solutions to enhance accuracy and efficiency.

\subsubsection{CKM Denoising}
In real-world measurement campaigns, noise inevitably corrupts the collected channel data due to sensor imperfections, environmental interference, and measurement errors, leading to inaccurate or distorted CKMs. To tackle this issue, the CKMImageNet dataset is used to train denoising models such as diffusion models\cite{luo2025denoising} and generative adversarial networks (GANs)\cite{salimans2016improved}. These models learn to filter out noise and recover the true channel characteristics by training on CKM data that has been deliberately corrupted with noise. By leveraging the spatially consistent and environment-aware features of CKMImageNet, these models can significantly improve the quality of CKMs, resulting in more accurate channel estimates for system optimization. 

\subsubsection{CKM Inpainting}
Another challenge in CKM construction arises when certain areas of the map are missing due to limitations such as high measurement or computational
costs, as well as the inability to conduct on-site measurements for safety-restricted or inaccessible areas. For example, if certain regions lack measurement data, traditional methods struggle to accurately infer the missing information. While the CKMImageNet dataset can be used to train inpainting models that restore these missing regions, enabling more complete data representation. A conditional decoupled diffusion model is trained in \cite{fu2024generative} using the CKMImageNet dataset. Furthermore, conditional diffusion models trained on CKMImageNet can inpaint large missing regions (e.g., entire city blocks) by leveraging environmental context from building layouts and material properties. Additionally, few-shot inpainting techniques enable rapid adaptation to novel environments using limited measurements, reducing reliance on exhaustive data collection.

\subsubsection{CKM Super-resolution}
Due to the high costs and equipment limitations, it is often challenging to obtain high-resolution CKMs in practice. Measurements may only be available at a low resolution, leading to a coarse representation of the channel. CKMImageNet helps address this problem by training super-resolution network, which learn to upscale low-resolution CKMs to a higher resolution. By using the high-quality, high-resolution grayscale images in CKMImageNet as ground truth, the network can enhance the spatial resolution of channel maps, providing finer details that are critical for tasks such as beamforming and interference management in 6G systems. In \cite{wang2024deep}, a ResNet is trained to perform super-resolution task. By training on paired low-resolution and high-resolution maps, models learn to recover fine details such as shadowing effects near building edges or subtle angle-of-arrival variations. This capability is critical for ultra-dense networks, where millimeter-level precision in beamforming requires sub-meter channel knowledge.

\subsubsection{Physical Environment Map to CKM Generation}
In some cases, channel knowledge data may be sparse or only a limited set of measurement points is available, while environmental maps (such as building layouts, terrain features, and urban infrastructure) might be accessible. Using CKMImageNet, environment-to-CKM mapping can be achieved. By training AI models on data that integrates environment maps with corresponding CKMs, the models learn to predict the full map based on sparse measurements and the environment maps. This approach allows for accurate channel estimation even when direct measurements are sparse, making it ideal for applications in complex environments where CKMs are not directly measurable.

\subsection{Case 2: CKM-Enhanced Channel Estimation}

In extremely large-scale multiple-input multiple-output (XL-MIMO) systems, challenges like high-dimensional channels and non-Gaussian distributions degrade traditional channel estimation methods. The maximum a posteriori (MAP) framework offers a promising solution by combining pilot measurements with channel priors, but its effectiveness relies on accurately modeling complex statistical distributions. CKMImageNet addresses this by leveraging historical channel data and deep denoising techniques to approximate high-precision priors. By integrating these learned priors into MAP-based optimization, CKMImageNet enables low-overhead, high-accuracy channel estimation, significantly outperforming conventional least squares (LS) and linear minimum mean square error (LMMSE) approaches in scenarios with limited pilots and non-Gaussian statistics. This enhancement improves robustness and efficiency in 6G environment-aware communication systems.

\subsection{Case 3: CKM-enhanced ISAC}

CKM strengthens sensing through capabilities like suppressing clutter and leveraging spatial relationships. For instance, it can filter out irrelevant reflections such as echoes from urban obstacles—using channel knowledge such as clutter angles, enabling the system to focus on target signals for better accuracy. Additionally, CKM inherently connects channel information to specific locations via multipath data, making it naturally suited for localization and sensing. This natural linkage supports precise detection and tracking in complex environments. CKMImageNet, as a dataset, boosts this further with visuals of channel and obstacle patterns. It helps AI grasp the intricate connections between channels and locations in diverse, challenging settings like isolating a vehicle's echo amid urban noise. By training on this data, AI refines its skill in decoding reflections and determining locations, enhancing sensing precision and efficiency for advanced applications.

\section{Conclusion And Futher Work}
This paper introduces CKMImageNet, a dataset designed to bridge the gap between AI and environment-aware wireless communication environment-aware research for 6G systems. CKMImageNet integrates location-specific channel data, multi-resolution channel knowledge images and physical environmental maps, with both numerical and visual representations. The dataset is constructed through physics-based ray-tracing simulations using advanced tools like Wireless Insite and Sionna, ensuring spatial consistency and environmental accuracy. By fusing channel parameters—such as channel gain, AoA, AoD with grayscale images and obstacle matrices, CKMImageNet enables AI models to learn environment-dependent propagation patterns while maintaining computational efficiency.
CKMImageNet serves as a foundational resource for a wide range of AI-driven tasks in wireless communications and sensing. It supports AI training for tasks like AI-based CKM construction (e.g., denoising, inpainting, and super-resolution) and advanced applications such as environment-aware channel estimation and sensing enhancement. 

Future developments of the CKMImageNet dataset will focus on expanding its scope and utility. This includes incorporating data from different frequency bands to better represent the diverse spectrum requirements of 6G and beyond. Additionally, the dataset will include data from various transmitter and receiver heights to account for the multi-level nature of modern communication infrastructures, such as those found in urban environments with multiple base stations at different heights. Furthermore, the dataset will be enriched with more dynamic environmental conditions, such as varying weather patterns and time-of-day effects, to enhance the realism and applicability of the data for AI-driven communication systems. These enhancements will make CKMImageNet an even more comprehensive and versatile tool for researchers and practitioners in the field of advanced wireless communications and AI integration.

\bibliographystyle{IEEEtran}
\bibliography{reference}

\end{document}